\documentstyle[12pt,aasms4,flushrt]{article}

\received{}
\accepted{}

\begin{document}

\title{Outburst in the Polarized Structure of the Compact Jet of 3C~454.3}

\author{Jos\'e-Luis G\'omez} 
\affil{Instituto de Astrof\'{\i}sica de Andaluc\'{\i}a, CSIC, Apartado 3004,
18080 Granada, Spain}
\authoremail{jlgomez@iaa.es}
\author{Alan P. Marscher}
\affil{Department of Astronomy, Boston University, 725 Commonwealth
Avenue, Boston, MA 02215, USA}
\authoremail{marscher@bu.edu}
\and
\author{Antonio Alberdi}
\affil{Instituto de Astrof\'{\i}sica de Andaluc\'{\i}a, CSIC,
Apartado 3004, 18080 Granada, Spain}
\authoremail{antxon@iaa.es}

\begin{abstract}

  We present three-epoch polarimetric images of the quasar 3C~454.3 obtained
with the Very Long Baseline Array at 22 and 43 GHz. 22 GHz polarized intensity
images show a sudden change in the polarization structure of a bright eastern
component (which we call the ``core,'' although it may be neither at the
upstream end of the jet nor completely stationary) over a 41-day interval,
coincident with the ejection of a new component from the core, as resolved in
the corresponding 43 GHz images. This polarization outburst is also present at
43 GHz in both the core and the new component. This may represent a rapid
change in the electric vector position angle of the ejected component from
being orthogonal to almost parallel to that of the core. About seven months
later, the new component, moving superluminally at 2.9$\pm$0.4 $h^{-1}\,c$
($q_{\circ}$=0.5) relative to the core and 3.9$\pm$0.4 $h^{-1}\,c$ relative to
a bright stationary component about 0.6 mas west of the core --- very low
compared with previous measurements --- is found at 43 GHz to exhibit a
further rotation of 90$^{\circ}$ in the orientation of its
polarization. Opacity effects may account for the first rotation, but changes
in the magnetic field of the component and/or that of the underlying jet in
the inner milliarcsecond structure of 3C~454.3 are needed to account for the
second.

  Polarized intensity images of the quasar 0420$-$014, used as a calibrator,
are also presented. The polarization position angle of the core rotated
between late 1994 and late 1996.

\end{abstract}

\keywords{Polarization - Techniques: interferometric - galaxies: active -
quasars: individual: 3C~454.3 - Galaxies: jets - Radio continuum: galaxies }

\section{Introduction}

  The quasar 3C~454.3 at redshift $z$=0.859 is one of the brightest
extragalactic radio sources. It is an optically violent variable with a
relatively high total linear polarization. It was the subject of the first
polarimetric VLBI observation (Cotton et al. \cite{Co84}) at a wavelength of
13 cm. Pauliny-Toth et al. (\cite{PT87}) presented the results of an extensive
monitoring program of this source, covering about 5 yr of observations at 2.8
cm. These observations revealed the existence of superluminal components, with
apparent proper motions between 0.21 and 0.35 mas yr$^{-1}$, or equivalently
between 4.4 and 7.3 $h^{-1}c$ ($H_{\circ}$= 100 $h$ km s$^{-1}$ Mpc $^{-1}$,
$q_{\circ}$=0.5), and a pair of quasi-stationary components, one situated at
about 0.6 mas of the core, and a second one at about 1 mas, the latter of
which was only detected from 1983.8 to 1984.9. A higher resolution
polarimetric 6 cm VLBI map was presented by Cawthorne \& Gabuzda
(\cite{CG96}). This showed a curving jet with a magnetic field aligned with
the jet axis, except for the inner component {\it K7} (which can be identified
with the stationary component at 0.6 mas observed by Pauliny-Toth et
al. \cite{PT87}), whose magnetic field lay almost perpendicular to the
direction of the jet flow. Kemball et al. (\cite{Ke96}) presented the first
polarimetric 7 mm Very Long Baseline Array (VLBA)\footnote{The VLBA is an
instrument of the National Radio Astronomy Observatory, which is a facility of
the National Science Foundation operated under cooperative agreement by
Associated Universities, Inc.} image of 3C~454.3. This revealed a
three-component structure in polarized intensity, consisting of the core, the
stationary component previously found by Pauliny-Toth et al. (\cite{PT87}) at
0.6 mas, and a new component located between both of these. This component was
determined to have a polarization position angle almost perpendicular to the
stationary component, which has a magnetic field perpendicular to the jet
axis, as previously observed by Cawthorne \& Gabuzda (\cite{CG96}). Marscher
(\cite{Al98}) presented a sequence of eleven 43 GHz VLBA images covering two
years of observations starting in late 1994. A proper motion of 0.28 $\pm$
0.02 mas yr$^{-1}$ was observed for a component in the inner milliarsecond
structure. These observations revealed no stationary component at the eastern
end of the jet, but rather a weak, roughly stationary component 0.2 mas
downstream of it. Further 5 and 8.4 GHz VLBI observations of 3C~454.3 were
presented by Pauliny-Toth (\cite{PT98}), showing mean proper motions of 0.68
$\pm$ 0.02 mas yr$^{-1}$ along a curved path at a distance of 2--4 mas from
the ``core.''

\section{Observations}

  We present three-epoch polarimetric observations of 3C~454.3 obtained with
the VLBA. The first two observations were performed on 1996 November 11 and
December 22 at 1.3 cm and 7 mm, in which 3C~454.3 served as a calibrator for
3C~120 (G\'omez et al. \cite{JL98}).  The data were recordered in 1-bit
sampling VLBA format with 32 MHz bandwidth per circular polarization. The
reduction of the data was performed within the NRAO Astronomical Image
Processing System (AIPS) software. The instrumental polarization was
determined using the feed solution algorithm developed by Lepp\"anen et
al. (\cite{Le95}) on the source 0420-014. Comparison with other polarimetric
observations carried out at similar epochs and frequencies revealed a good
agreement in the determination of the D-terms (M. Lister, private
communication), and were used to calibrate the electric-vector position angle
(EVPA) with an error that we estimate to be within 10$^{\circ}$. We also refer
the readers to G\'omez et al. (\cite{JL98}) for further details about the
reduction and calibration of the data.

  Figure \ref{0420} shows the VLBA images obtained for 0420-014. Only slight
structural changes are observed between both frequencies and epochs. The total
and polarized intensity images are dominated by a strong core component, with
an EVPA in the east-west direction, which smoothly rotates toward the
southwest, in the direction of the jet structure. This represents a rotation
of the EVPA by about 45$^{\circ}$ with respect to that measured by Kemball et
al. (\cite{Ke96}) two years previously, accompanied by a decrease in the peak
percentage linear polarization to less than 2.2\%.

  The third 43 GHz polarimetric VLBA observation took place on 1997 July
30--31 as part of a $\gamma$-ray blazar monitoring program. The data analysis
was similar to the previous two epochs, except that a component of 3C~279,
with its EVPA exactly parallel to the component's position angle measured with
respect to the core, was used to calibrate the EVPA. We estimate an error in
the EVPA absolute orientation to be less than 5$^{\circ}$ for this epoch.

\section{Results}

  Figures \ref{m3c454} and \ref{3c454} show VLBA images of 3C~454.3 obtained
with the VLBA at 22 and 43 GHz at the three epochs. Tables \ref{13mmfit} and
\ref{7mmfit} summarize the physical parameters obtained for
3C~454.3. Tabulated data correspond to total flux density ($S$), polarized
flux density ($p$), degree of polarization ($m$), EVPA ($\chi$), separation
($r$) and structural position angle ($\theta$) relative to the easternmost
bright component [which we refer to as the ``core'' despite the fact that this
component is not always bright (Marscher \cite{Al98}) and may not be
completely stationary (see below)], and angular size (FWHM).  Components in
the total intensity images were analyzed by model fitting the uv data with
circular Gaussian components within the software Difmap (Shepherd, Pearson, \&
Taylor \cite{Se94}). Components seen in the images of polarized intensity are
not always coincident with maxima in total intensity (see also Kemball et
al. \cite{Ke96}; G\'omez et al. \cite{JL98}), therefore we can only obtain
estimates of $m$ by the approximation that both maxima are coincident.

\subsection{Source structure}

  Total intensity 1.3 cm images show a double component structure, very
similar to that observed by Kemball et al. (\cite{Ke96}) at 7 mm, consisting
of the core and component {\it St}. This appears at a very similar separation
from the core to that of component {\it 2} in Pauliny-Toth et
al. (\cite{PT87}), component {\it K7} of Cawthorne \& Gabuzda (\cite{CG96}),
and component {\it 2} of Kemball et al. (\cite{Ke96}). Hence, we identify it
as the same stationary component, first detected in 1983.8 by Pauliny-Toth et
al. (\cite{PT87}). We notice, however, that {\it St} seems to have been
observed at different position angles from the core, from values close to
$-100^{\circ}$ measured by Pauliny-Toth et al. (\cite{PT87}) and Cawthorne \&
Gabuzda (\cite{CG96}), to about $-70^{\circ}$ measured by Kemball et
al. (\cite{Ke96}), in closer agreement to the values we obtain. This is
indicative of a swing toward the north in the inner jet between the 1980's and
1990's. Furthermore, our images show a systematic difference between the
position angles and distances from the core of {\it St} for both observing
wavelengths. This may be due to changes in the internal structure of {\it St}
or the core --- most probably the latter, since such changes are expected
during the ejection of new components from the core. Indeed, our 7 mm images
reveal the existence of component {\it A}, blended with the core in the
corresponding 1.3 cm images.

  Figure \ref{3c454} also shows a component between the core and {\it St}, at
a distance from the core in good accordance with that expected from
extrapolating the motion of {\it A} between the previous two epochs (see
Fig. \ref{m3c454}).  Therefore, we identify it as the same component,
obtaining a proper motion relative to the core for the three combined epochs
of $\mu$=0.14 $\pm$0.02 mas yr$^{-1}$, which corresponds to an apparent speed
of 2.9$\pm$0.4 $h^{-1}\,c$. Relative to component {\it St}, the proper motion
is $\mu$=0.18$\pm$0.02 mas yr$^{-1}$, or 3.9$\pm$0.4 $h^{-1}\,c$. In either
case, the motion is significantly slower than that observed by Marscher
(\cite{Al98}), who measured $\mu$=0.28 $\pm$0.02 mas yr$^{-1}$ for a component
found at 43 GHz to be moving between the core and {\it St} during the period
December 1994-October 1995. Unless the orientation with respect to the
observer of the inner jet in 3C~454.3 changed significantly between 1995 and
1996-1997, this would imply the ejection of components with intrinsically very
different velocities in order to account for the differences in the observed
proper motions. Even faster velocities were detected at larger scales by
Pauliny-Toth (\cite{PT98}), whose observations at 5 and 8.4 GHz between 1984
and 1991.9 indicated a component moving along a curved path with a mean proper
motion of $\mu$=0.68 $\pm$0.02 mas yr$^{-1}$.

  Assuming that {\it A} has maintained a constant speed since its ejection
from the core, we estimate its birth at approximately 1995.7, coincident with
a small outburst detected by the University of Michigan monitoring at 14.5
GHz\footnote{The University of Michigan Radio Astronomy Observatory is
supported by the National Science Foundation and by funds from the University
of Michigan.} and Mets\"ahovi Radio Research Station monitoring data at 22 and
37 GHz (Ter\"asranta et al. \cite{Hi98}).  Kemball et al. (\cite{Ke96})
observed a component --- only detected in polarization --- at a very similar
separation from the core to that of {\it A} in July 1997, most probably
associated with a previous ejection which they estimate took place at about
1994.4.

  While component {\it A} does not show significant changes in flux during the
three epochs, the core experienced a significant decrease in flux by about 1.5
Jy between the 1996.98 and 1997.58 epochs. Component {\it St} shows a very
similar flux between both of the 1.3 cm epochs, however at 7 mm its flux
progressively decreased, with a total variation of more than 2 Jy between the
1996.86 and 1997.58 epochs. Similar large variations of flux were also found
in {\it St} by Pauliny-Toth et al. (\cite{PT87}) at 2.8 cm.

  Beyond component {\it St}, Fig. \ref{3c454} shows a complex jet structure,
with emission extending to the north and south. Model fitting reveals a faint
component in the north direction, which we have labeled {\it n}; it appears in
the polarized intensity images as well. A lower resolution 15 GHz image
presented by Kellermann et al. (\cite{Ke98}) shows some indications of this
structure in the form of a very extended core emission. 3C~454.3 is observed
to extend initially to the west, presenting a relatively strong bend toward
the northwest direction at about 4-5 mas from the core (Pauliny-Toth et al.
\cite{PT87}, Pauliny-Toth \cite{PT98}; Cawthorne \& Gabuzda \cite{CG96};
Kellermann et al. \cite{Ke98}).

  Figure \ref{3c454} shows a faint extension of emission east of the core
position. Unless it is due to calibration errors, its presence indicates
emission upstream of the ``core.''  A similar structure was first detected by
Marscher (\cite{Al98}) in a series of 43 GHz VLBA images covering about two
years, starting December 1994. Other indications of emission upstream of the
core have also been found in high resolution 43 GHz VLBA images of 3C~120
(G\'omez et al. \cite{JL98}). A possible interpretation is that it corresponds
to the actual region where the jet is being generated; such weak emission
could be due to a lower flow Lorentz factor, as in the accelerating jet model
of Marscher (\cite{Al80}), or associated with the birth of new moving
components upstream of the core (Marscher \cite{Al98}). In this case the core
would represent a recollimation shock --- strong in our observations and weak
in 1995 when observed by Marscher (\cite{Al98}), as expected theoretically
(Daly \& Marscher \cite{DM88}; G\'omez et al. \cite{JL95}).

\subsection{Polarization}

  Polarized intensity images corresponding to the epochs at the end of 1996,
shown in Fig. \ref{m3c454}, reveal a sudden change in the polarized structure
of the core at both observing frequencies in a 41-day interval. At 1.3 cm the
core changes from being almost undetected in polarization, to showing a
polarized flux of 78 mJy, with a degree of polarization of $m$=2.2\%. This
change in polarized flux is accompanied by a rotation of $\chi$ by almost
60$^{\circ}$.  At 7 mm we observe a very similar situation, in which the core
and component {\it A} remain undetected in polarization at the 1996.86 epoch,
but at 1996.98 appear with a degree of polarization for the core similar to
that observed at 1.3 cm, and a dramatic increase in polarization for {\it A},
with $m$=7.4\%. Both the core and {\it A} show a similar $\chi$ to that
observed at 1.3 cm for the core. The third 7 mm epoch reveals a rotation in
$\chi$ of about 105$^{\circ}$ for component A, accompanied by a small decrease
in $m$, while the core remained with a similar $\chi$ but a reduced percentage
polarization to values similar to that corresponding to 1996.86 at 1.3 cm.

  Component {\it St} is detected at all observing epochs and frequencies. Its
degree of polarization maintains values between 1.3 and 2.2\%, similar to that
observed at 5 GHz by Cawthorne \& Gabuzda (\cite{CG96}), except for the July
1997 epoch, in which it increases to 8.6\%, close to that obtained by Kemball
et al. (\cite{Ke96}). Component {\it St} seems to have a frequency dependent
EVPA. Our 7 mm images show $\chi$ close to the east-west direction, similar to
that observed by Kemball et al. (\cite{Ke96}). However, at 1.3 cm {\it }the
EVPA of {\it St} presents a systematic offset of about 30--40$^{\circ}$ with
respect to the values measured at 7 mm. Cawthorne \& Gabuzda (\cite{CG96})
obtained a value of 29$^{\circ}$ in observations at 6 cm, which suggest a
rotation of $\chi$ in {\it St} toward the north-south direction with
increasing wavelength. Broten, Macleod, \& Vall\'ee (\cite{BMV88}) obtained a
rotation measure of -57 rad m$^{-2}$ for 3C~454.3, which may account for the
rotation of $\chi$ between our 1.3 cm values and those presented by Cawthorne
\& Gabuzda (\cite{CG96}). However, this small rotation measure would not
affect our 1.3 cm and 7 mm observations, and consequently no Faraday rotation
corrections have been made for the measured EVPA.

\section{Theoretical interpretation and conclusions}

\subsection{Stationary component {\it St}}

  The existence of stationary and moving components in 3C~454.3 is also found
in several other sources, e.g. in 4C~39.25 (Alberdi et al. \cite{An93}). In
this case, the stationary component is explained as being produced by a bend
toward the observer, with the increased flux due to enhanced Doppler
boosting. Similarly, we could explain the stationarity of {\it St} as due to a
bend toward the line of sight. Indeed, the jet of 3C~454.3 is observed to bend
toward the northwest direction at about 4-5 mas from the core (e.g.,
Pauliny-Toth et al. \cite{PT87}, Pauliny-Toth \cite{PT98}).  In this case, a
change in the apparent motion of component {\it A} is expected as it moves
along the hypothetical bent trajectory, as observed and simulated in the case
of 4C~39.25. Computing the proper motion of {\it A} between each pair of
consecutive epochs, we find some evidence of deceleration as it moves closer
to {\it St}, as well as a progressive, although small change in its position
angle toward the northwest. With the information provided by the proper motion
of {\it A} derived from the three epochs combined, we can estimate a maximum
viewing angle of about 38$^{\circ}$, and a minimum Lorentz factor of 3.07. In
order to obtain a deceleration of {\it A} as produced by a bend toward the
observer, the viewing angle must be significantly smaller than the maximum
allowed, which final value depends on the actual Lorentz factor of {\it A}. In
this case component {\it A} should also experience an increase in its flux,
due to an enhancement of its Doppler boosting. However, only minor changes in
the flux of {\it A} are observed across the three epochs. Of course, it is
also possible that the apparent small deceleration of {\it A} could also
result from a true change in its bulk Lorentz factor. However, the large error
affecting the proper motion determination between the first two epochs
prevents from drawing any solid conclusion regarding a possible deceleration
of {\it A}.

  The existence of a bend toward the observer in the region of {\it St} could
also explain the large changes in flux experienced by this component, with no
significant changes in its position and structure. A moving component should
pass through {\it St}, giving during the interaction the impression of a
single component with increasing flux, subsequently fading progressively as it
passes {\it St} and turns around the bend (e.g., G\'omez et
al. \cite{JL94}). Unless the the jet in {\it St} bends in a plane containing
the observer, which is a priori unlikely, future components that pass through
{\it St} should move in a different direction in the plane of sky after the
event.

  Another possibility to explain the stationarity of {\it St} is that it is
produced by a recollimation shock in the jet flow. Numerical simulations of
the relativistic hydrodynamics and emission of jets have shown that pressure
mismatches between the jet and the external medium may result in the
generation of internal oblique shocks (G\'omez et al. \cite{JL95},
\cite{JL97}). These shocks appear in the emission as stationary components due
to the increased specific internal energy and rest-mass density. When a moving
component passes through one of these recollimation shocks, both components
would blend to appear as a single feature. This is accompanied by a
``dragging'' of the merged components downstream, because of the increase in
the Mach number, as well as an enhancement of the emission. After the
collision, the two components would split up, with the previously stationary
component associated with the recollimation shock progressively fading and
recovering its initial position and flux. The would give the appearance of
motion upstream, as long as the initial physical conditions in the jet are
recovered (G\'omez et al. \cite{JL97}). Within this scenario, a moving
component would not experience significant changes in its proper motion and
flux as it approached the stationary component, similar to what is observed
for {\it A}. This would give the impression of a quiescent merge of the two
components. A similar situation, accompanied by a brief dragging of the
stationary component, was observed for the merging of components {\it K1} and
{\it K2} by Gabuzda et al. (\cite{De94}) in the BL Lac object 0735+178. In the
case of a strong standing shock, as perhaps produced by a sudden change in the
external medium pressure, more violent interactions with moving shocks may be
expected.

  We propose a consistent scenario for the inner region in 3C~454.3 in which
both the core and component {\it St} represent strong recollimation
shocks. When new components are generated, they should increase significantly
the emission of the core, briefly dragging its position downstream. This
interpretation is in very good agreement with the observations by Marscher
(\cite{Al98}). These show a roughly stationary component, labeled S2
(corresponding to the component marked as the core in this paper) at about 0.2
mas downstream of the eastern end of the jet. New superluminal components seem
to appear upstream of S2, which could explain the emission upstream of the
core in Fig. \ref{3c454}. These observations also showed a slight motion
downstream of S2, before it recovered its initial position as a moving
component passes it, as predicted by the theory (G\'omez et
al. \cite{JL97}). A similar interaction would be expected when the moving
component reaches the position of the next recollimation shock, corresponding
to {\it St}. The large changes in the flux of {\it St} could then be explained
by these interactions. However, in order to test this model, accurate
measurements of the absolute positions of components are needed, possible
through a careful high resolution phase-reference monitoring program. They
should provide the necessary information to confirm the constancy of the
core--{\it St} separation (within the expected motions due to the passage of
moving components) and the emergence and evolution of new components upstream
of the core.

  Cawthorne \& Cobb (\cite{CC90}) showed that, depending on the jet flow
Lorentz factor and the orientation with respect to the observer, conical
shocks may show a polarization position angle parallel or perpendicular to the
jet flow.  In the limit of strong shocks, this would require small viewing
angles, as measured in the rest frame of the shock, to explain the aligned
EVPA with respect to the projected direction of the jet observed for the core
and {\it St}.

  Tables \ref{13mmfit} and \ref{7mmfit} reveal an optically thin spectrum for
component {\it St} at both 1996 epochs, which eliminates opacity effects as
being responsible for the systematic offset observed in the EVPA at different
wavelengths. The observed EVPA for component {\it St} at 1.3 cm differs only
by about 20$^{\circ}$ from that measured by Cawthorne \& Gabuzda (\cite{CG96})
at 6 cm. Some of this discrepancy may be explained by Faraday rotation, which
at 6 cm would produce a rotation of $\sim -12^{\circ}$ (Broten, Macleod, \&
Vall\'ee \cite{BMV88}). Hence, the measured difference in the EVPA these
authors observed between {\it St} and the outer components is probably due to
an intrinsically different nature, rather than to opacity effects. A
possibility is that the outer components Cawthorne \& Gabuzda (\cite{CG96})
observed represent weak plane-perpendicular moving shocks in a predominantly
longitudinal magnetic field configuration. These shocks will slightly enhance
the perpendicular component of the magnetic field (parallel to the shock
front), but the increase will not be enough to overcome the initial
longitudinal field. Hence the final net orientation of the field would remain
aligned in the direction of the jet flow. As a consequence of the partial
cancellation of the magnetic field produced by the shocks, a small degree of
polarization is expected, which contrasts with the large values measured by
Cawthorne \& Gabuzda (\cite{CG96}) for the outer components. This cannot be
applied to {\it St}, and we need to consider a different interpretation in
terms of a conical shock or a bend, as outlined previously. If {\it St}
corresponds to a bend, no changes in the EVPA are expected as a consequence of
the change in curvature along the bent portion of the jet, and we need to
assume an underlying perpendicular magnetic field for {\it St} to explain the
observed EVPA, as opposed to that measured downstream by Cawthorne \& Gabuzda
(\cite{CG96}). Another possibility, still under the hypothesis of a bend along
the position of {\it St}, is that there is another moving plane perpendicular
shock component passing through it, such that {\it St} represents the blended
component, whose parallel EVPA is due to the enhancement by the shock of the
perpendicular component of the magnetic field. However, it seems unlikely to
have such a situation each time the source has been observed. In the case that
{\it St} corresponds to an oblique shock, the observed EVPA can still be
explained without needing to assume a change in the underlying magnetic field
of {\it St} with respect to the outer components, in which case the magnetic
field would be aligned with the jet axis throughout the entire jet. Depending
on the jet flow Lorentz factor and viewing angle, Cawthorne \& Cobb
(\cite{CC90}) showed that a conical shock may exhibit an EVPA aligned with the
jet axis. Those results were obtained considering an initially randomly
oriented magnetic field, and need to be confirmed in the case of an initially
aligned field.

\subsection{Polarized outburst in superluminal component {\it A}}

  The polarized structural outburst observed in the 1996 epochs may be
explained by assuming that, at the 1996.86 epoch, the EVPA of the core and
component {\it A} were mutually perpendicular, producing a net cancellation of
the polarized intensity. In this case, component {\it A} is required to have
changed its EVPA by almost a full rotation of 90$^{\circ}$, making it
approximately aligned with that of the core -- assumed to remain with an
approximately constant EVPA --- which led to the sudden appearance of both
components in the polarized intensity images at 1996.98. Taking into account
that the apparent core at 1.3 cm in fact corresponded to blending of the core
and component {\it A}, the resulting spectrum is rather flat, and we could
assume that the rotation of 90$^{\circ}$ in the EVPA of {\it A} may be due to
a change from being optically thick at 1996.86, to optically thin at
1996.98. We shall also note that the lack of polarization in the core region
at the first epoch may be due to large opacity values.  However, the fact that
{\it A} remained undetected at 22 GHz prevents us from obtaining a reliable
determination of its spectrum, and hence its opacity. It is also possible that
the burst in polarization may result from drastic changes in the magnetic
field configuration.

    This represents a remarkably rapid change in the polarized structure of
component {\it A}. From the timescale of variability, we can derive an upper
limit to the size of {\it A} as $R_{\rm max}=c\,t_{{\rm
var}}\,[\delta/(1+z)]$, where $\delta$ is the Doppler factor. Using the
minimum Lorentz factor of 3.07 derived from the proper motion, and assuming a
viewing angle of 1/$\Gamma$, which maximizes the apparent velocity, we obtain
a maximum size of $\sim$10 $\mu$as. Model fitting component sizes tabulated in
table \ref{7mmfit} show that only for epoch 1997.58 can the size of component
{\it A} be measured, with an estimated FWHM of 190 $\mu$as, well above the
derived maximum. We shall note however, that we cannot rule out that the
observed variability may arise from the core. In this case the estimated sizes
tabulated in Table \ref{7mmfit} are very similar and also above the estimated
maximum, except for epoch 1996.86 in which model fitting yields a
delta-function component for the core. We therefore conclude that the Doppler
factor must be higher than the minimum implied by these observations, and
instead must be similar to those suggested by the faster proper motions
observed by Marscher (\cite{Al98}) and Pauliny-Toth (\cite{PT98}).

  If component {\it A} changed its opacity from optically thick to thin
between 1996.86 and 1996.98, it seems less plausible to assume that {\it A}
became thick again in 1997.58, as would be required if we were to explain the
further rotation of 90$^{\circ}$ in its EVPA in this way.  To account for this
extra rotation, we need to assume a change in the magnetic field of the
underlying jet or in component {\it A}. If component {\it A} is associated
with a moving plane-perpendicular shock, we expect an EVPA aligned with the
direction of the jet flow when the component is optically thin. This could
explain the value of $ \chi $ observed in 1996.98. Since the underlying
magnetic field remains aligned with the jet axis through the jet of 3C~454.3,
as seems to be deduced from the outer components observed by Cawthorne \&
Gabuzda (\cite{CG96}), the rotation of 90$^{\circ}$ in the EVPA of {\it A}
could be explained by assuming that the strength of the shock associated with
it decreases as the component moves downstream. In 1996.98 the enhancement of
the perpendicular component of the magnetic field produced by the shock
associated with {\it A} would overcome the initial longitudinal field
direction. However, once the shock has moved to the position observed in
1997.58, we find that the enhancement of the perpendicular field by a weaker
shock would not be enough to change the initially aligned net field of the
underlying jet, resulting in a net magnetic field parallel, and an EVPA
perpendicular, to the jet axis. However, it then remains unclear why component
{\it A} maintained a similar flux --- even experiencing a small increase, as
opposed to what it would be expected in the case of a decrease in the shock
strength. Within this scenario, component {\it St} would be required to be
associated with a conical shock in order to obtain a net magnetic field
perpendicular to the jet axis (assuming no magnetic field changes between the
positions of {\it A} and {\it St}).

  Further polarimetric high-resolution VLBA observations are required to test
these hypotheses. The study would be improved significantly by performing
phase-reference to an external source, allowing a detailed determination of
the proper motions of components. These would be of great importance to test
our hypothesis of recollimation shocks to interpret the nature of {\it St} and
possibly the core. In this case, numerical simulations predict a temporary
drag of these components, followed by a brief upstream motion to recover their
initial positions (G\'omez et al. \cite{JL97}). Polarimetric observations
would provide the necessary information to discern between a possible rotation
of the underlying magnetic field configuration, or a change in the strength of
shocks in the inner structure of 3C~454.3.

\begin{acknowledgements}
  This research was supported in part by Spain's Direcci\'on General de
Investigaci\'on Cient\'{\i}fica y T\'ecnica (DGICYT), grants PB94-1275 and
PB97-1164, by NATO travel grant SA.5-2-03 (CRG/961228), by U.S. National
Science Foundation grant AST-9802941, and by NASA through CGRO Guest
Investigator Program grants NAG5-2508, NAG5-3829, and NAG5-7323 and RXTE Guest
Investigator Program grants NAG5-3291, NAG5-4245, and NAG5-7338.
\end{acknowledgements}

\newpage

\begin{figure}
\figcaption{VLBA images of 0420-014 at 22 GHz at epochs 1996 November 11 ({\it
top left}) and 1996 December 22 ({\it top right}), and 43 GHz at 1996
November 11 ({\it bottom left}) and 1996 December 22 ({\it bottom
right}). Total intensity is plotted as contour maps, while the linear gray
scale shows the linearly polarized intensity. The superposed sticks give the
direction of the electric field vector.  From {\it top left} to {\it bottom
right} images: contour levels are in factors of 2, starting at [of peak
intensity] 0.5 [3.15], 0.2 [6.5], 0.25 [3.45] and 0.2\% [1.7 Jy beam$^{-1}$];
convolving beam [position angle]: 0.67$\times$0.32 [-1$^{\circ}$],
0.69$\times$0.31 [-1$^{\circ}$], 0.39$\times$0.17 [-2$^{\circ}$] and
0.39$\times$0.17 (FWHM) [-2$^{\circ}$] mas; maximum in polarized intensity
[noise level]: 51 [5], 78 [23], 49 [10] and 37 [7] mJy
beam$^{-1}$.\label{0420}}
\end{figure}

\begin{figure}
\figcaption{Same as Fig. \ref{0420} but for 3C~454.3. From {\it top left} to
{\it bottom right} images: contour levels are in factors of 2, starting at [of
peak intensity] 3 [2.49], 2 [3.03], 2 [2.1] and 2\% [2.45 Jy beam$^{-1}$];
convolving beam [position angle]: 0.75$\times$0.27 [-1$^{\circ}$],
0.79$\times$0.31 [-2$^{\circ}$], 0.39$\times$0.14 [-2$^{\circ}$] and
0.39$\times$0.15 (FWHM) [-2$^{\circ}$] mas; maximum in polarized intensity
[noise level]: 58 [20], 84 [9], 33 [6] and 83 [8] mJy
beam$^{-1}$.\label{m3c454}}
\end{figure}

\begin{figure}
\figcaption{Same as Fig. \ref{0420} for 3C~454.3, epoch 1997 July 30 at 43
GHz. Contour levels are in factors of 2, starting at 0.5\% (including -0.5\%)
of the peak of 0.94 mJy beam$^{-1}$. The convolving beam is 0.342$\times$0.151
mas (FWHM), with position angle -1$^{\circ}$, and the maximum in polarized
intensity is 69 mJy beam$^{-1}$, with a noise level of 6.9 mJy
beam$^{-1}$. \label{3c454}}
\end{figure}

\newpage

\begin{deluxetable}{lccccccc}
\tablecolumns{8}
\tablewidth{0pc}
\tablecaption{22 GHz models for 3C~454.3.\label{13mmfit}}
\tablehead{
 & $S$ & $p$ & $m$ & $\chi$ & $r$ & $\theta$ & FWHM\nl
Component & (Jy) & (mJy) & (\%) & ($^{\circ}$) & (mas) & ($^{\circ}$) &(mas)\nl
}
\startdata
\multicolumn{8}{c}{1996.86}\nl
\hline 
Core\dotfill & 3.11 & 26 & 0.8 & 178 & ... & ... & 0.23\nl
St\dotfill & 4.54 & 59 & 1.3 & 50 & 0.51 & -67 & 0.37\nl
\hline 
\multicolumn{8}{c}{1996.98}\nl
\hline
Core\dotfill & 3.59 & 78 & 2.2 & 120 & ... & ... & 0.21\nl
St\dotfill & 4.61 & 86 & 1.8 & 47 & 0.53 & -66 & 0.38\nl
\enddata
\end{deluxetable} 

\begin{deluxetable}{lccccccc}
\tablecolumns{8}
\tablewidth{0pc}
\tablecaption{43 GHz models for 3C~454.3.\label{7mmfit}}
\tablehead{
& $S$ & $p$ & $m$ & $\chi$ & $r$ & $\theta$ & FWHM\nl
Component & (Jy) & (mJy) & (\%) & ($^{\circ}$) & (mas) & ($^{\circ}$) &(mas)\nl
}
\startdata
\multicolumn{8}{c}{1996.86}\nl
\hline 
Core\dotfill & 2.09 & ... & ... & ... & ... & ... & 0\nl
A\dotfill & 1.48 & ... & ... & ... & 0.16 & -93 & 0\nl
St\dotfill & 3.59 & 78 & 2.2 & 81 & 0.62 & -72 & 0.39\nl
\hline
\multicolumn{8}{c}{1996.98}\nl
\hline 
Core\dotfill & 2.53 & 55 & 2.2 & 96 & ... & ... & 0.03\nl
A\dotfill & 1.16 & 86 & 7.4 & 125 & 0.18 & -87 &0\nl
St\dotfill & 2.45 & 34 & 1.4 & 89 & 0.61 & -71 &0.38\nl
\hline 
\multicolumn{8}{c}{ 1997.58}\nl
\hline 
Core\dotfill & 0.92 & 9 & 0.9 & 109 & ... & ... & 0.03\nl
A\dotfill & 1.28 & 68 & 5.2 & 20 & 0.26 & -85 & 0.19\nl
St\dotfill & 1.50 & 100 & 8.6 & 93 & 0.59 & -70 & 0.42\nl
n\dotfill & 0.13 & 1.5 & 4.3 & 10 & 1.13 & -39 & 0.18\nl
\enddata
\end{deluxetable} 

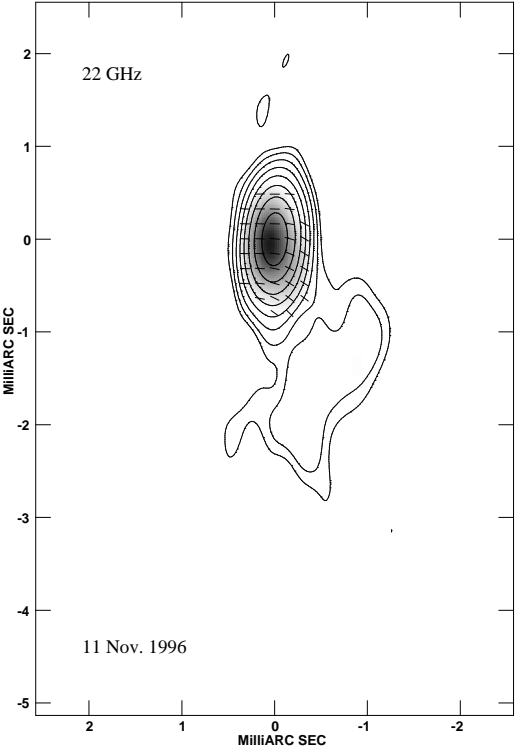
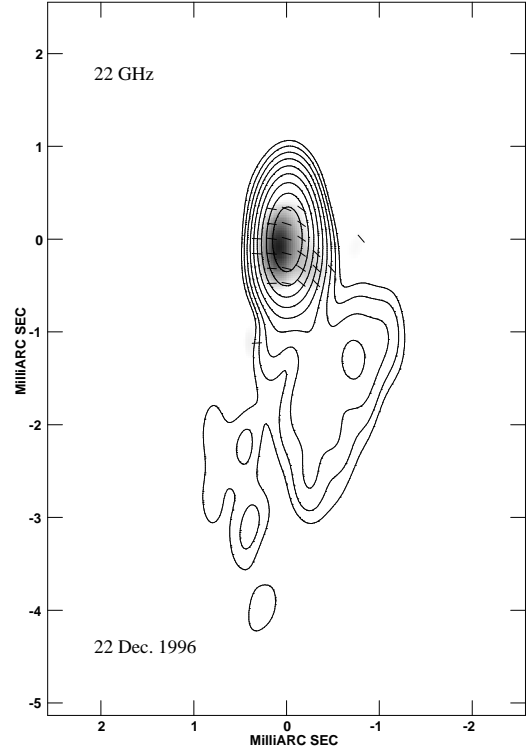
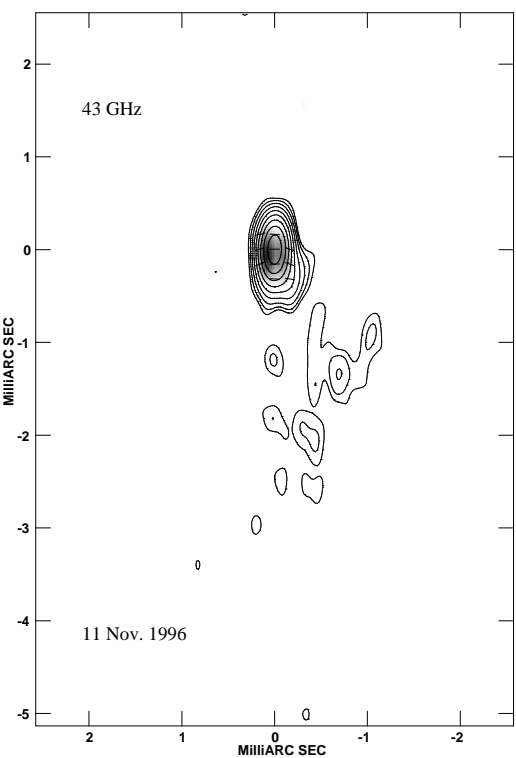
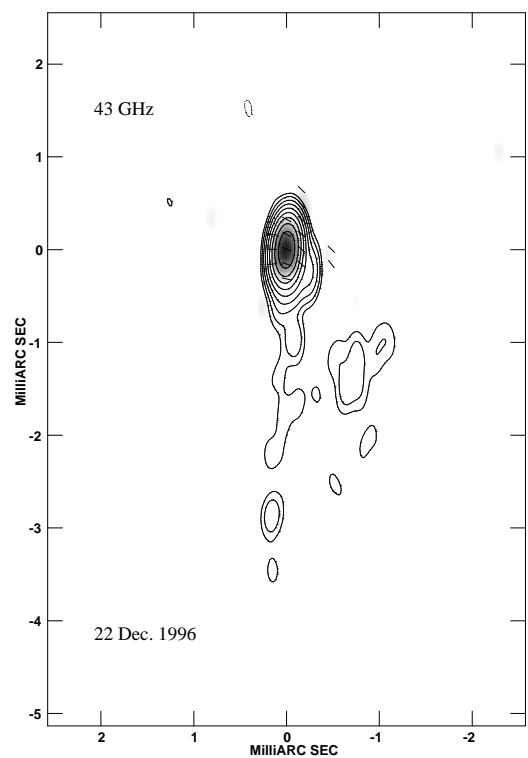
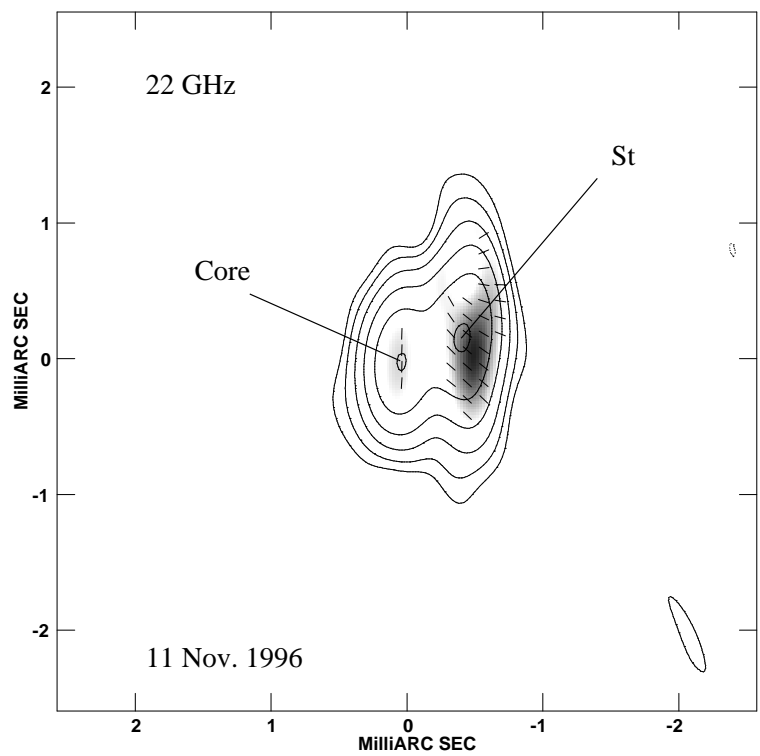
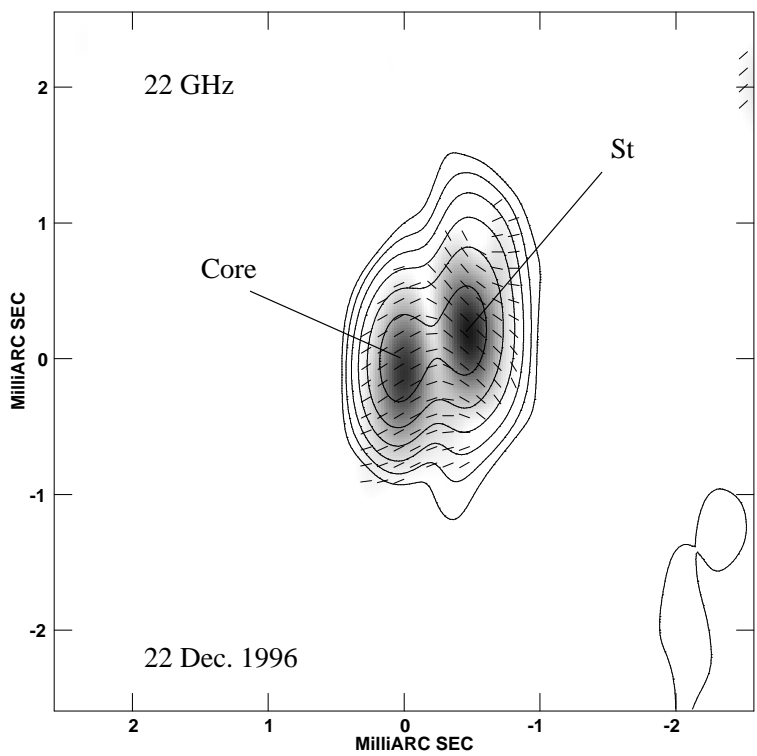
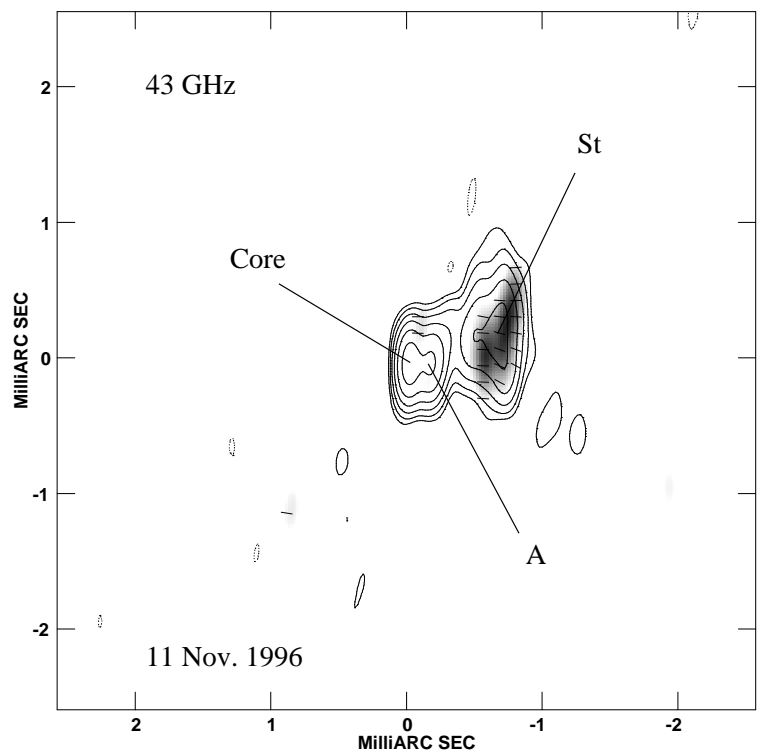
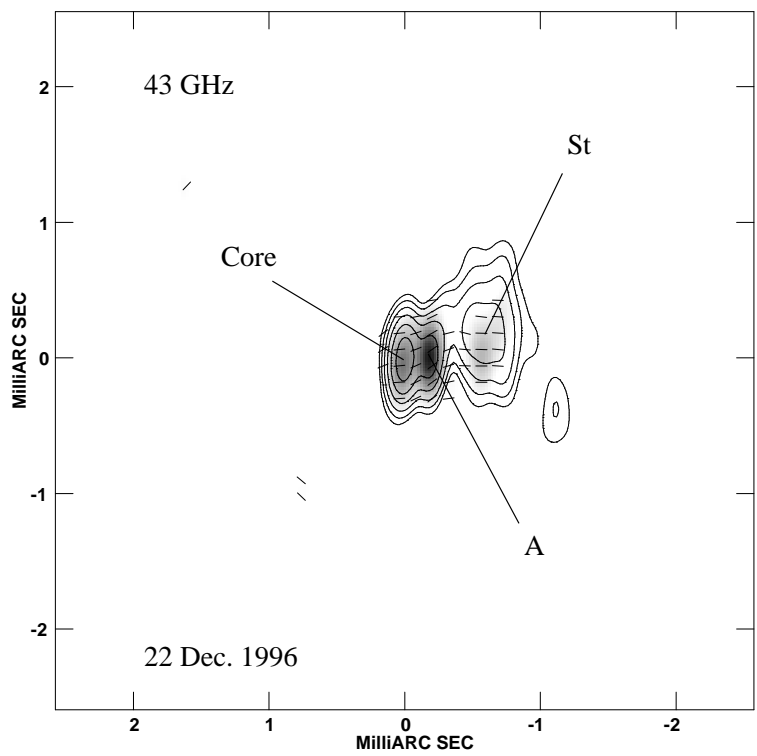
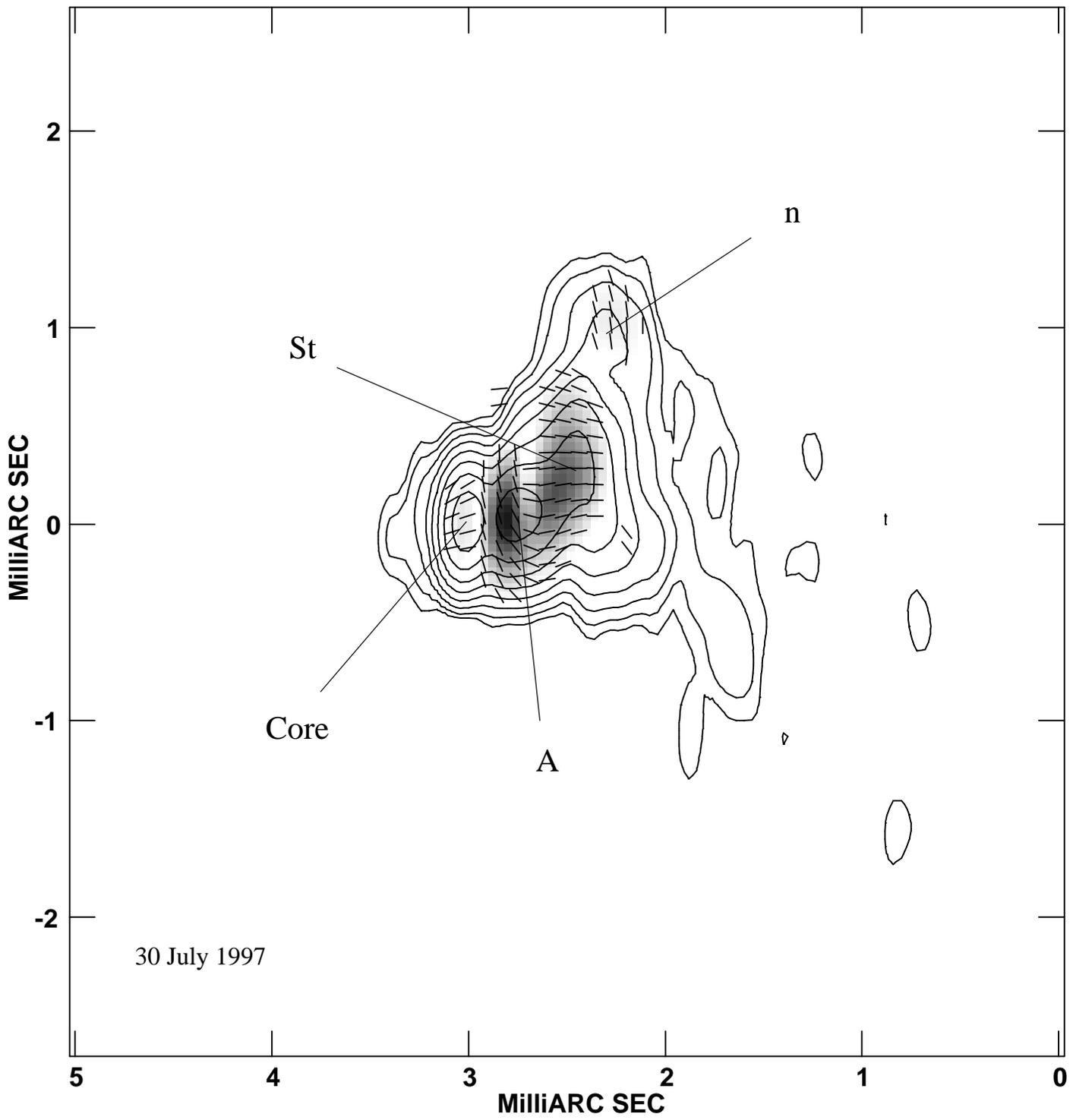
\end{document}